\documentclass{ws-mpla}

\def\be{\begin{equation}}
\def\ee{\end{equation}}
\def\ba{\begin{eqnarray}}
\def\ea{\end{eqnarray}}

\newcommand{\fr}[2]{\frac{#1}{#2}}
\def\ga{\mathrel{\raise.3ex\hbox{$>$\kern-.75em\lower1ex\hbox{$\sim$}}}}
\def\la{\mathrel{\raise.3ex\hbox{$<$\kern-.75em\lower1ex\hbox{$\sim$}}}}

\begin{document}

\markboth{Seokcheon Lee} {Time Varying Fine Structure Constant and
Proton-Electron Mass Ratio with Quintessence}

%%%%%%%%%%%%%%%%%%%%% Publisher's Area please ignore %%%%%%%%%%%%%%
\catchline{}{}{}{}{}
%%%%%%%%%%%%%%%%%%%%%%%%%%%%%%%%%%%%%%%%%%%%%%%%%%%%%%%%%%%%%%%%%%%

\title{Time Variation of Fine Structure Constant and
Proton-Electron Mass Ratio with Quintessence}

\author{\footnotesize Seokcheon Lee}

\address{Institute of Physics, Academia Sinica, \\ Taipei, Taiwan, 11529, R.O.C.\\
skylee@phys.sinica.edu.tw}

\maketitle

\pub{Received (31 January 2007)}{Revised (Day Month Year)}

\begin{abstract}
Recent astrophysical observations of quasar absorption systems
indicate that the fine structure constant $\alpha$ and the
proton-electron mass ratio $\mu$ may have evolved through the
history of the universe. Motivated by these observations, we
consider the cosmological evolution of a quintessence-like scalar
field $\phi$ coupled to gauge fields and matter which leads to
effective modifications of the coupling constants and particle
masses over time. We show that a class of models where the scalar
field potential $V(\phi)$ and the couplings to matter $B(\phi)$
admit common extremum in $\phi$ naturally explains constraints on
variations of both the fine structure constant and the
proton-electron mass ratio.

\keywords{Time Varying Fine Structure Constant; Proton-Electron
Mass Ratio; Quintessence.}
\end{abstract}

\ccode{PACS Nos.: include PACS Nos.}

\section{Introduction}

Recent observations show non-vanishing time variation of the fine
structure constant $\alpha$ by use of the relativistic shifts of
atomic energy levels in quasar absorption spectra\cite{Webb} and
nontrivial time evolution of the proton-electron mass ratio $\mu =
m_{p}/m_{e}$ from the observations of $H_{2}$ spectral lines in
quasars \cite{pemass}.

However there have been attempts to detect a variation in $\alpha$
using similar methods \cite{Petitjean} that has shown null results
so far. This rather controversial situation becomes even more
complicated if other constraints on $\Delta \alpha / \alpha $ are
taken into account \cite{Uzan}. The Oklo natural reactor and
meteoritic abundances of rhenium provide stringent constraints on
the change of the coupling constants that goes back to $z\sim 0.1
- 0.4$ \cite{Oklo,OPQ,fuj}. However, a recent re-analysis of Oklo
phenomenon suggested that the present data are consistent with the
non-zero change of $\Delta\alpha/\alpha = 4.5 \times 10^{-8}$
\cite{Lam}.

In spite of its questionable status of the non-zero claim for
$\Delta \alpha/\alpha$ there have been a number of attempts
\cite{Chiba,Wett,AG,CNP,Bek,Damour,LS,LV,OP,SBM} to build simple
models that could account for a possible $O(10^{-5})$ relative
shift in the fine structure constant at redshifts $z\sim 1$. The
Bekenstein model \cite{Bek} with a scalar field coupled to both
the electromagnetic field and dark matter \cite{Damour,OP,SBM} can
drive the time evolutions of fine structure constant and masses.

In quintessence models, which is one of the most commonly proposed
candidates for dark energy\cite{Ratra,FJ} to explain the current
accelerating universe\cite{SCP} we can naturally have the change
of the coupling constant\cite{CK,BMS,CWett,LLN,NL,AMO,LOP,MR,AMNO}
and masses of particles\cite{JBarr,CKYY} over cosmological times
due to the interaction of quintessence field with gauge fields and
matter.

In this paper we consider the scalar field in the Bekenstein model
as a quintessence field to check the time variations of the fine
structure constant and the proton-electron mass ratio. As the
extension of our previous paper\cite{LOP}, we briefly show the
result of the cosmological evolution of the dark energy and dark
matter energy density over the redshift. We show that the
cosmology of coupled quintessence models with a common extremum in
$V(\varphi)$ and gauge and matter/gauge couplings $B_i(\varphi)$
which can be consistent with all observational requirements. This
paper is organized as follows. In the next section we introduce
our model and display the necessary cosmological equations. In
sections 3, we make predictions for the variation of the coupling
constants. We consider the time varying masses in section4.
Section 5 is devoted to the conclusions.

\section{Cosmological Evolution of the Coupled Scalar Field}

The action including the interaction of a dimensionless light
scalar field $\phi$ with matter and gauge fields is given by, \ba
S_{\phi} &=& \int d^4x \sqrt{-g} \Biggl\{ \frac{\bar{M}^2}{2}
[\partial^{\mu} \phi \partial_{\mu} \phi - R] - V(\phi) -
\frac{B_{Fi}(\phi)}{4}F^{(i)}_{\mu\nu}F^{(i)\mu\nu} \nonumber
\\ && + \sum_j [\bar\psi_j iD\!\!\!\!/ \psi_j -
B_j(\phi)m_j\bar\psi_j\psi_j]\Biggr\}, \label{lagrangian} \ea
where $\bar M = M_{p}/\sqrt{8\pi}$ is the reduced Planck mass,
$B_{Fi}(\phi)$ represents the $\phi$-dependence of the gauge
couplings in Standard Model where the sum is over all three
groups. $\psi_j$ represents both Standard Model fermions and other
matter field ({\em i.e.} scalar Higgses, Majorana neutrinos etc.).
Since $\phi$ couples to the trace $T_\mu^\mu$ of dark matter, our
results are independent of the nature of the dark matter particles
(scalars or fermions).

Given a potential $V(\phi)$, the perfect fluid energy density and
pressure contributions due to $\phi$ are: \ba
\rho_{\phi} &=& \frac{1}{2} \bar{M}^2 \dot{\phi}^2 + V(\phi) \label{rho} \\
p_{\phi} &=& \frac{1}{2} \bar{M}^2 \dot{\phi}^2 - V(\phi) \equiv
\omega_{\phi} \rho_{\phi} \label{p}, \ea where the parameter
$\omega_\phi$ is the equation of state (EOS) of the scalar field.
By including the scalar field, we have the following Friedmann
equation in a Robertson-Walker Universe, \ba H^2 &=& \frac{1}{3
\bar{M}^2} \Bigl( \rho_{\phi} + \rho_{r} + \rho_{m} \Bigr) \equiv
\frac{1}{3 \bar{M}^2} \rho_{cr} \label{H}, \ea where $H=\dot a/a $
is the Hubble expansion rate, $\rho_{r}$ and $\rho_{m}$ is the
energy density of radiation and matter, and $\rho_{cr}$ is the
critical energy density. Notice the $\phi$-dependence of the
energy density of matter, {\em i.e.} $\rho_{m} =
\sum_{j}B_{j}(\phi)m_{j} <\bar{\psi}_{j} \psi_{j}>  \equiv
\sum_{j}B_{j}(\phi)m_{j} n_{j}$. Using these definitions, we can
write the scalar field equation, \be \ddot{\phi} + 3 H \dot{\phi}
+ \frac{1}{\bar{M}^2} \frac{\partial V}{\partial \phi} = -
\frac{1}{\bar{M}^2} \frac{\partial \ln B_{m}}{\partial \phi}
\rho_{m} . \label{eqphi} \ee

For cosmological studies that span a large range of redshifts $z$,
it is convenient to introduce the variable $x$ as the logarithm of
the scale factor $a$, \be x= \ln a = - \ln (1 + z) \label{x} \ee
where we choose the present scale factor $a^{(0)} = 1$.  With the
use of the variable $x$, we can rewrite the relevant system of
equations for $d\ln \rho_i/dx$ in the following form, \ba \frac{d
\ln \rho_{m}}{d x} &=& - 3 ( 1 + \omega_{m} ) + \frac{\partial \ln
B_{m}(\phi)}{\partial \phi} \frac{d \phi}{d x},
\label{rhoi1} \\
\frac{d \ln \rho_{r}}{d x} &=& - 3 ( 1 + \omega_{r} ),
\label{rhor1}
\\
\frac{d \ln \rho_{\phi}}{d x } &=& -3 ( 1 + \omega_{\phi} ) -
\frac{\partial \ln B_{m}(\phi)}{\partial \phi} \frac{d \phi}{d x}
\frac{\rho_{m}}{\rho_{\phi}}, \label{rhophi1} \ea where
$\omega_{r} = 1/3$ and $\omega_{m} = 0$ should be used for
radiation and matter respectively.

The derivative of $\phi$ with respect to $x$ can be obtained from
Eq (\ref{rho}), \be \left(\frac{d \phi}{d x}\right)^2 = 3
\Omega_{\phi} ( 1 + \omega_{\phi} ), \label{dphi} \ee If
$\omega_{\phi}$ is close to $-1$, the kinetic energy of the scalar
field goes to zero which occurs when the scalar field is close to
the minimum of the potential.

When the change in  $B_m(\phi)$ is not small and cannot be
neglected, the scaling of the matter energy density differs from
usual $a^{-3}$ behavior because of the changing mass, due to
$B_m(\phi)$, while the scaling of radiation energy density remains
unchanged, \ba \rho_{m}(x) &=& \rho_{m}^{(0)} a^{-3}
\frac{B_{m}\Bigl(\phi(x) \Bigr)}{B_{m} \Bigl(\phi(0)
\Bigr)},~~~{\rm and} ~~~ \rho_{r}(x) = \rho_{r}^{(0)} a^{-4}
\label{rhor2} \label{rhoi2}
\\ \rho_{r} &=& \frac{a_{eq}}{a} \frac{B_{m}\Bigl(\phi(x_{eq})
\Bigr)}{B_{m} \Bigl(\phi(x) \Bigr)} \rho_{m} \ea With the use of
these relations, we find \ba \frac{d \ln(1 -\Omega_{\phi})}{d x}
&=& \Omega_{\phi} \Biggl[3 \omega_{\phi} -
\Bigl(\frac{a_{eq}^{(c)}}{a+ a_{eq}^{(c)}} \Bigr) \Biggr] +
\Bigl(\frac{a}{a+ a_{eq}^{(c)}} \Bigr) \frac{\partial \ln
B_{m}}{\partial \phi} \frac{d \phi}{dx} \label{dOmega1} \\ \frac{d
\ln(1 - \omega_{\phi})}{d x} &=&  3( 1 + \omega_{\phi} ) +
\frac{\partial \ln V}{\partial \phi} \frac{d \phi}{dx} + \frac{(1
- \Omega_{\phi})}{\Omega_{\phi}} \Bigl(\frac{a}{a+ a_{eq}^{(c)}}
\Bigr) \frac{\partial \ln B_{m}}{\partial \phi} \frac{d \phi}{dx}
\label{domega1} \ea
where we have introduced an auxiliary
function, $a_{eq}^{(c)}(\phi) = a_{eq} B_{m}\Bigl(\phi(x_{eq})
\Bigr)B^{-1}_{m} \Bigl(\phi(x) \Bigr)$.

\begin{figure}
\vspace{1.8cm} \centerline{\psfig{file=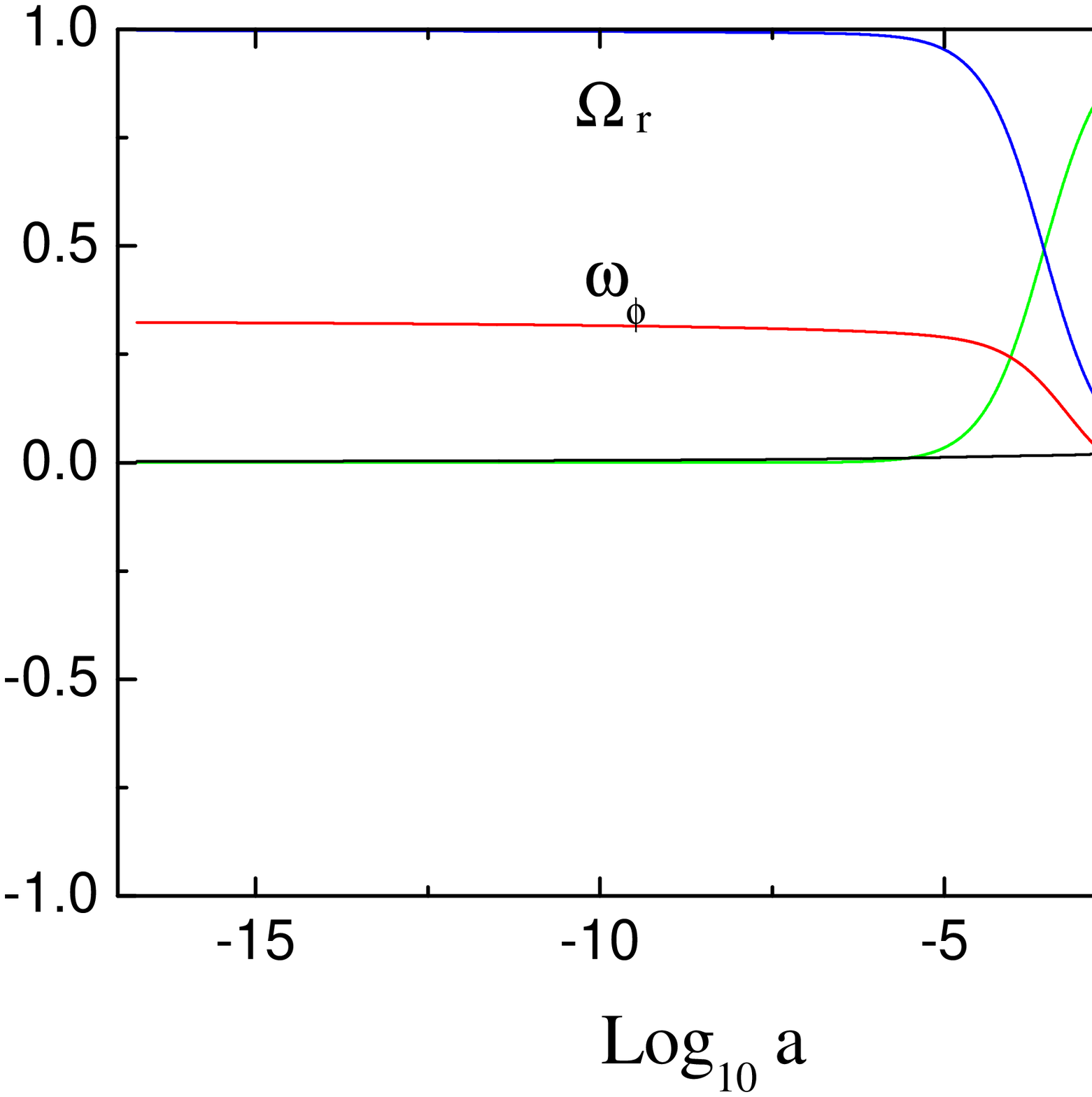,width=8cm}
\psfig{file=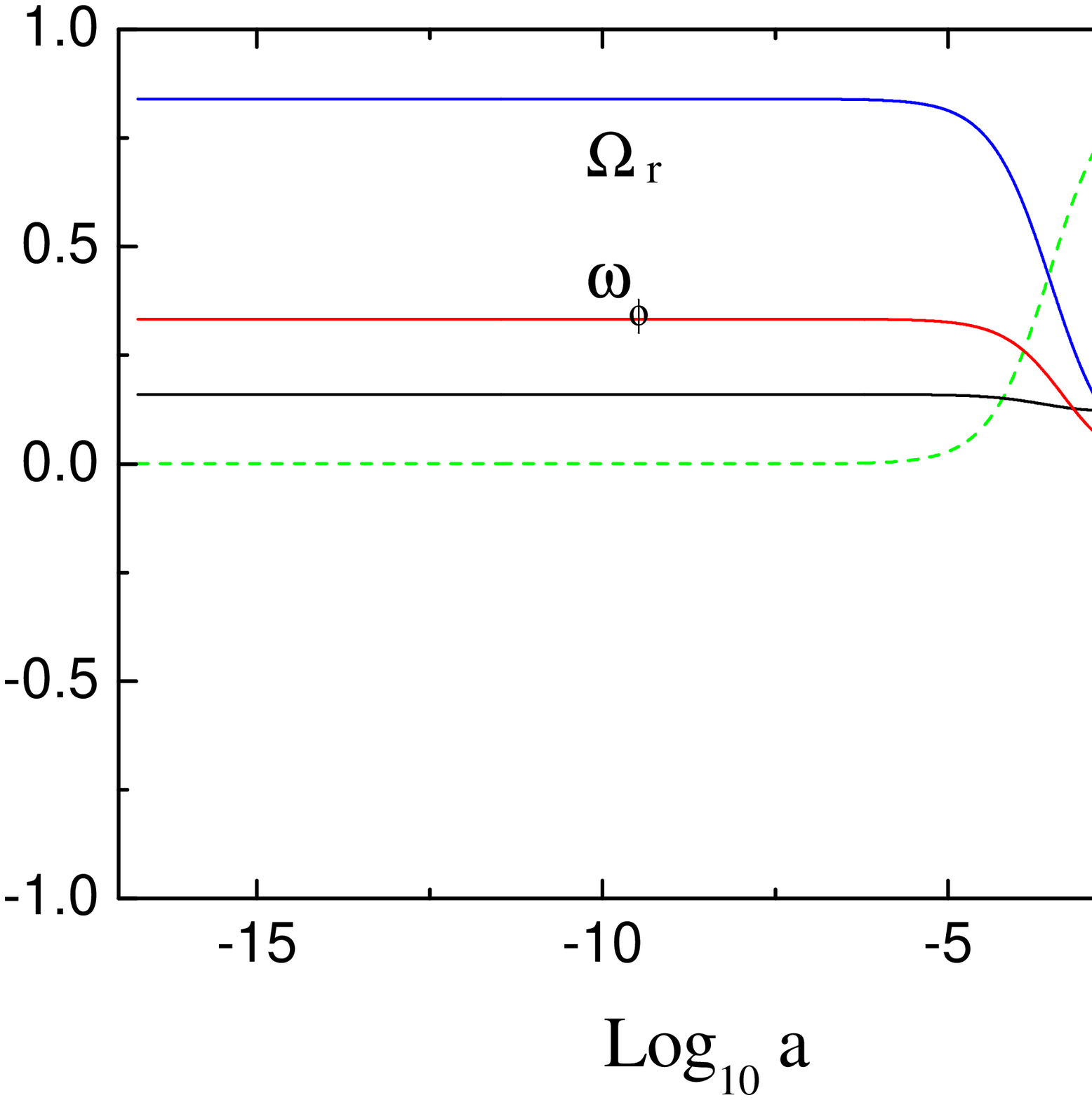,width=8cm} } \vspace{-1.7cm} \caption{The
cosmological evolution of the equation of state parameter,
$\omega_{\phi}$, and the energy density parameters, $\Omega_{i}$,
of each component for $\lambda = 5$ when $n = 10^{-3}$. a) When
the potential is $V= V_0 \exp(\lambda\phi^2/2)$. b) For the $V =
V_0 \cosh(\lambda\phi)$ potential. \protect\label{fig1}}
\end{figure}

We assume that all functions $B_i(\phi)$ and $V(\phi)$ admit a
common extremum, which is a generalization of Damour-Nordtvedt and
Damour-Polyakov constructions \cite{DN,DP}. Near this extremum,
all functions admit an expansion \be B_i(\phi) = 1 + \zeta_{i}
\phi + \fr{1}{2}\xi_i \phi^2+...;\;\;\;\;\;\;\; V(\phi) = V_0( 1 +
\lambda \phi + \fr{1}{2}\lambda \phi^2+...), \label{dp} \ee where
$\zeta$, $\xi_i$ and $\lambda$ are dimensionless parameters, while
$V_0$ is of the order of the dark energy density today.

We consider two simple potentials $V(\phi)$ and follow the same
ansatz for functions $B_i(\phi)$ as in our previous work
\cite{LOP}. \ba \label{Vphi}
{\rm case ~~A}: ~~~~~~~V(\phi) &=& V_{0} \exp \Bigl( \frac{\lambda\phi ^2 }{2} \Bigr)\\
{\rm case ~~B}: ~~~~~~~V(\phi) &=& V_{0} \cosh(\lambda\phi)
\label{Vphi1} \ea \be B_{i}(\phi) =
\Biggl(\frac{b_i+V(\phi)/V_0}{1+b_i}\Biggr)^{n_i}, ~~~ {\rm
where}~~ b_i+1>0 \label{BF2} \ee

We show the cosmological evolution of the equation of state
parameter, $\omega_{\phi}$, and the energy density parameters,
$\Omega_{i}$, of each component for both potentials in
Fig.~\ref{fig1}. In this figure we use $\lambda = 5$ and $n =
10^{-3}$ which is a reasonable value adopted from previous work
\cite{LLN2}.

%%%%%%%%%%%
\begin{center}
\begin{figure}
\vspace{1.8cm} \epsfxsize=5cm \centerline{\psfig{file=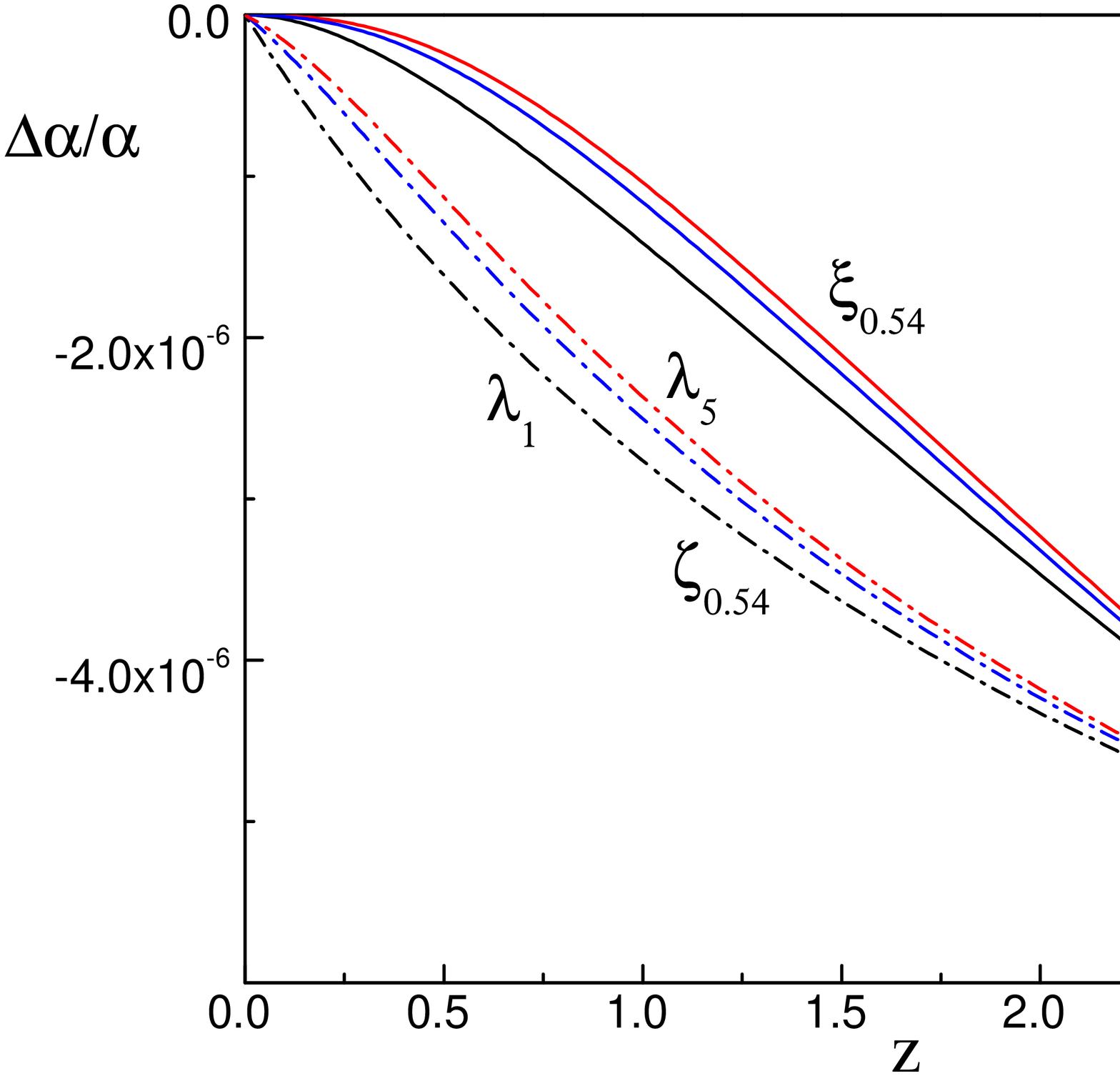,
height=4.5cm}\psfig{file=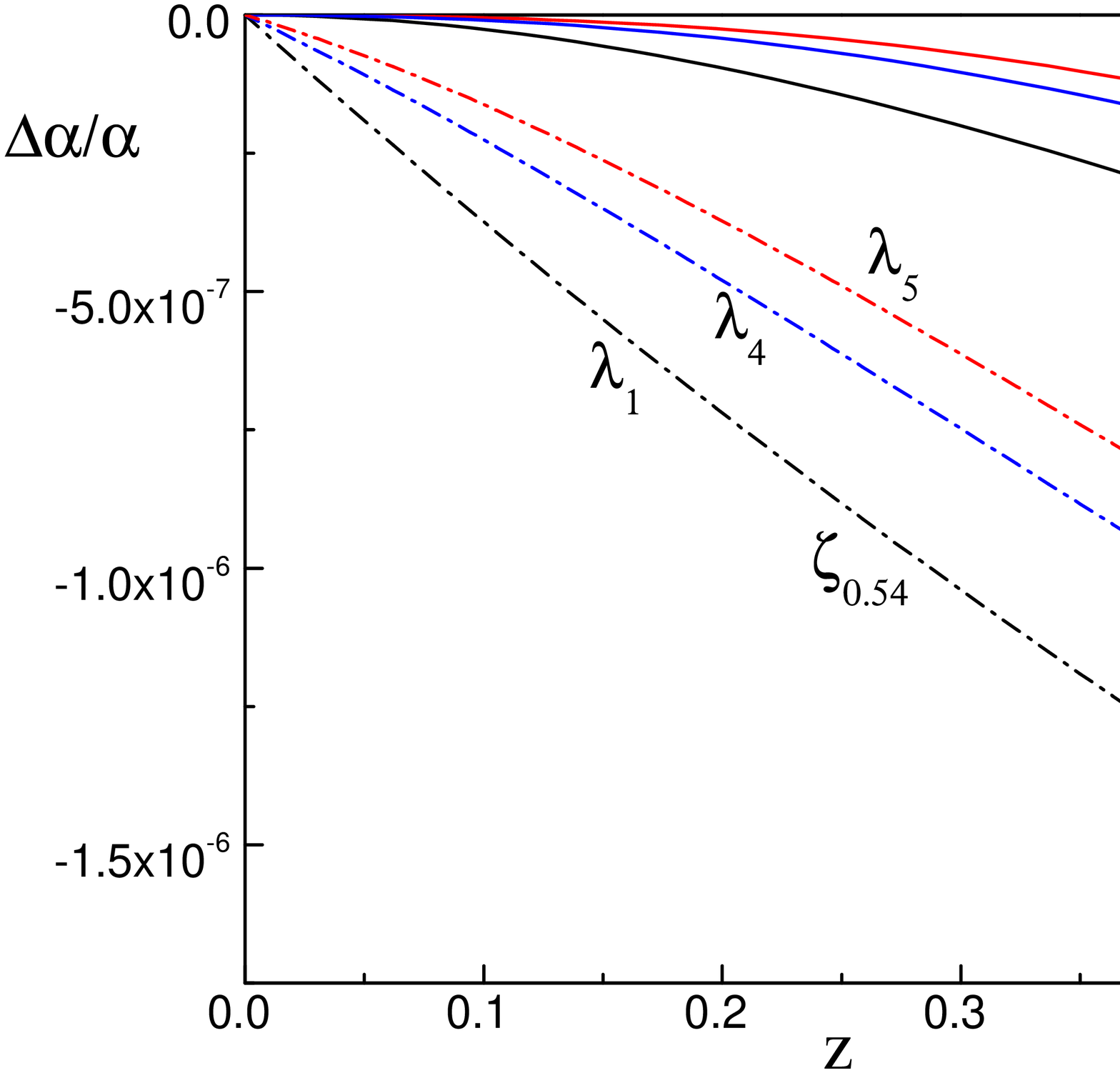, height=4.5cm}}
\epsfxsize=5cm \centerline{\psfig{file=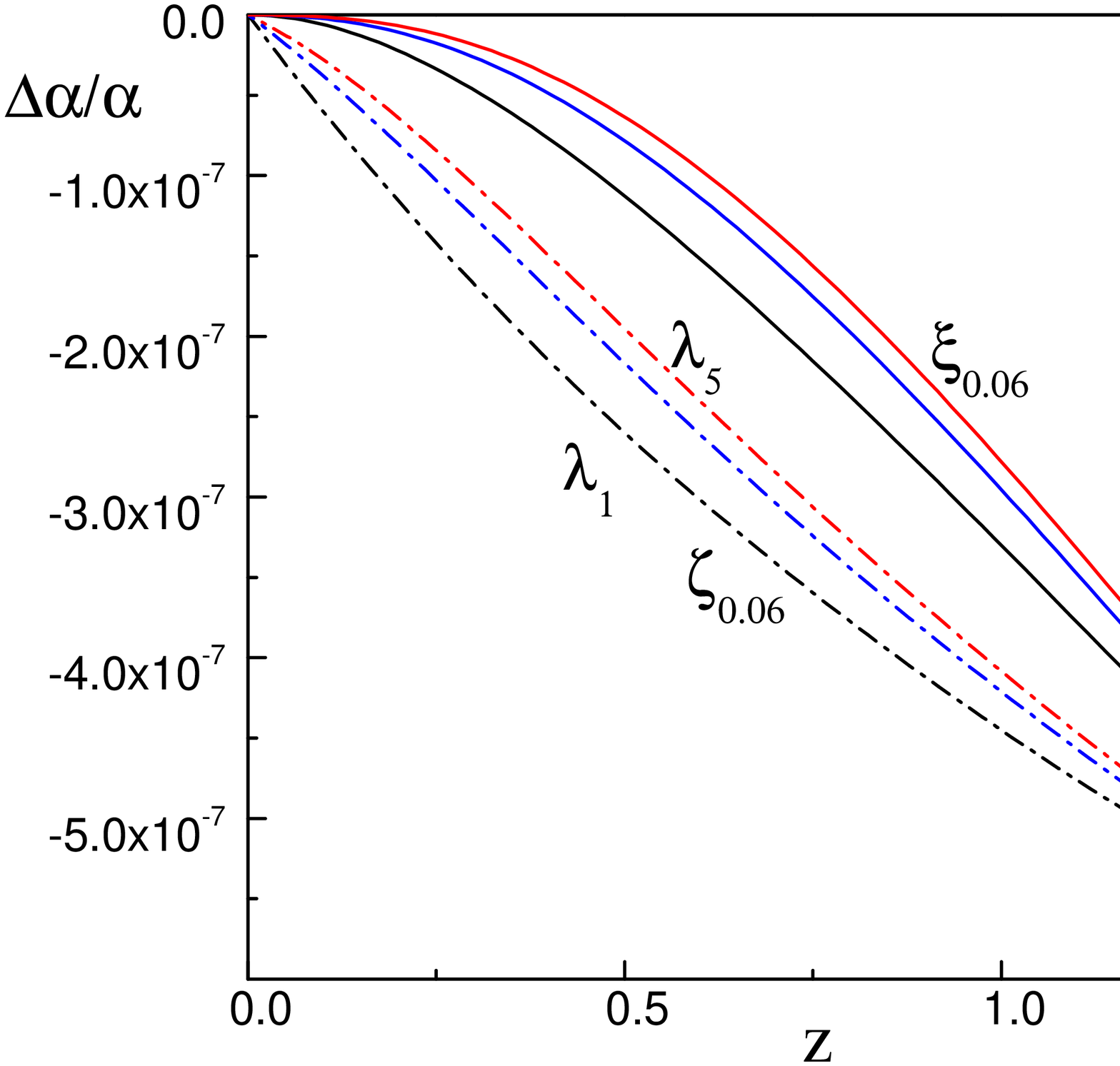,
height=4.5cm}\psfig{file=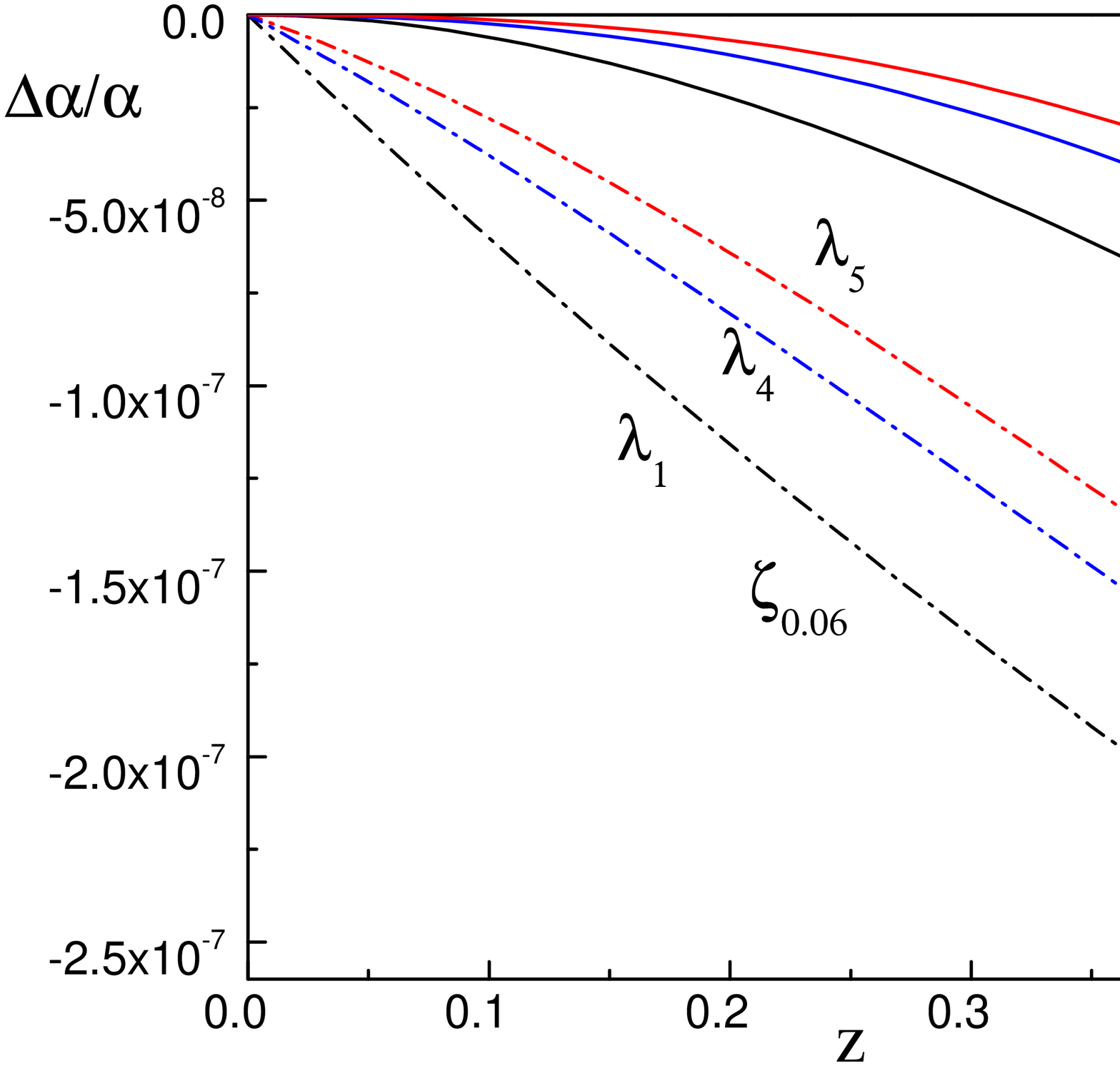, height=4.5cm}}
\vspace{-1.7cm} \caption{ a. The evolution of
$\Delta\alpha/\alpha$ over the redshift range $0\leq z\leq 3$
driven by the potential (\ref{Vphi}). Panels a) and b) use the
common normalization $\Delta \alpha/\alpha = -0.54 \times 10^{-5}$
at $z=3$. Figures c) and d) use the common normalization $\Delta
\alpha/\alpha = -0.06 \times 10^{-5}$ at $z=1.5$. The solid lines
correspond to the choice $\zeta = 0$ and $\xi \ne 0$, whereas the
dashed-dotted lines allow $\zeta \ne 0$. } \label{fig2}
\end{figure}
\end{center}
\vspace{-0.8cm}
\section{Time Variation of the Fine Structure Constant}

To study the cosmological evolution of the fine structure constant
we use the following relation between $B_F$ and $\alpha$, \be
\fr{\Delta \alpha(z)}{\alpha} \equiv
\fr{\alpha(z)-\alpha(0)}{\alpha(0)} =
\fr{B_F(\phi(0))}{B_F(\phi(z))}-1. \label{BF} \ee Note that the
present value of the field $\phi(0)$ is close to zero.

\begin{center}
\begin{figure}
\vspace{1.8cm} \epsfxsize=5cm
\centerline{\psfig{file=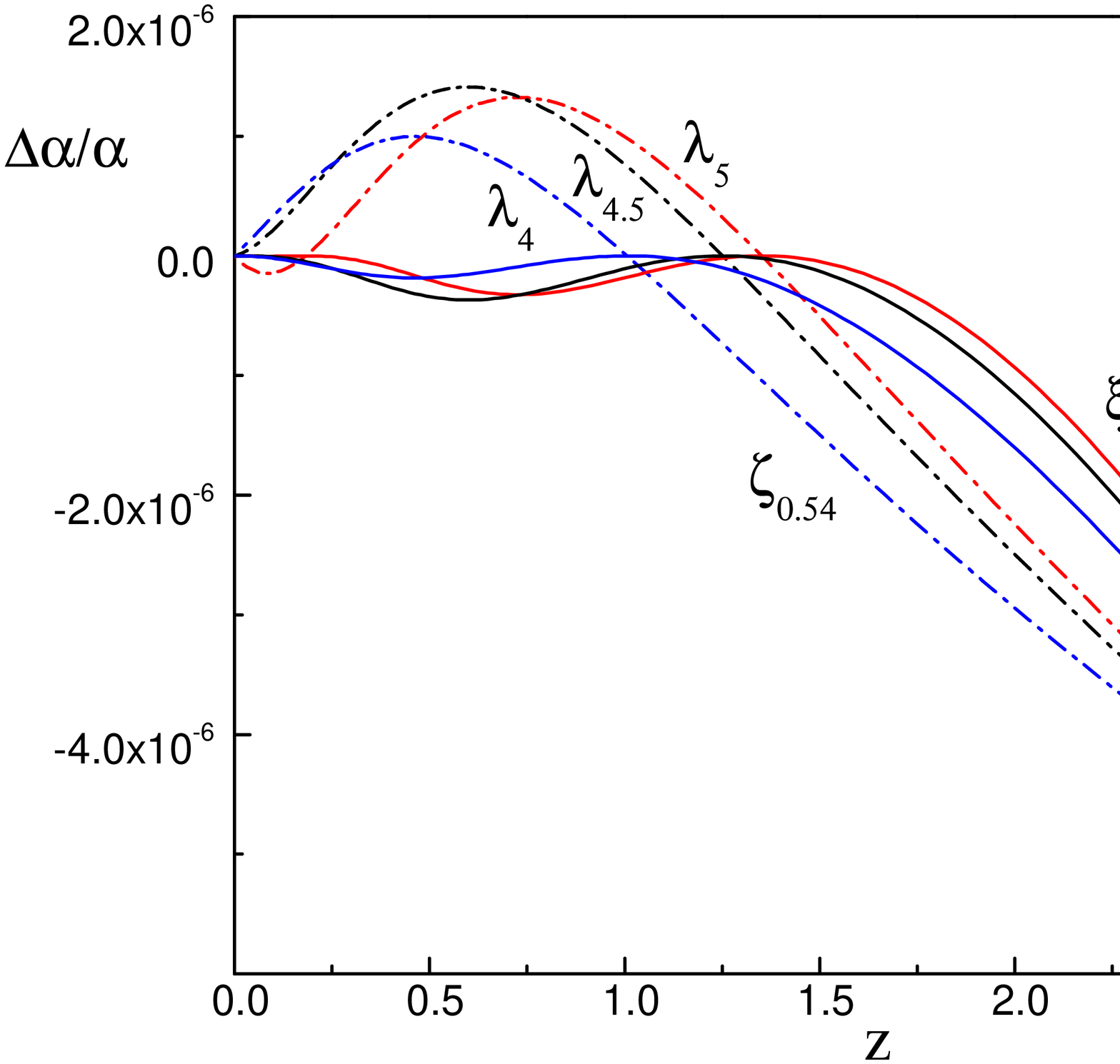,
height=4.5cm}\psfig{file=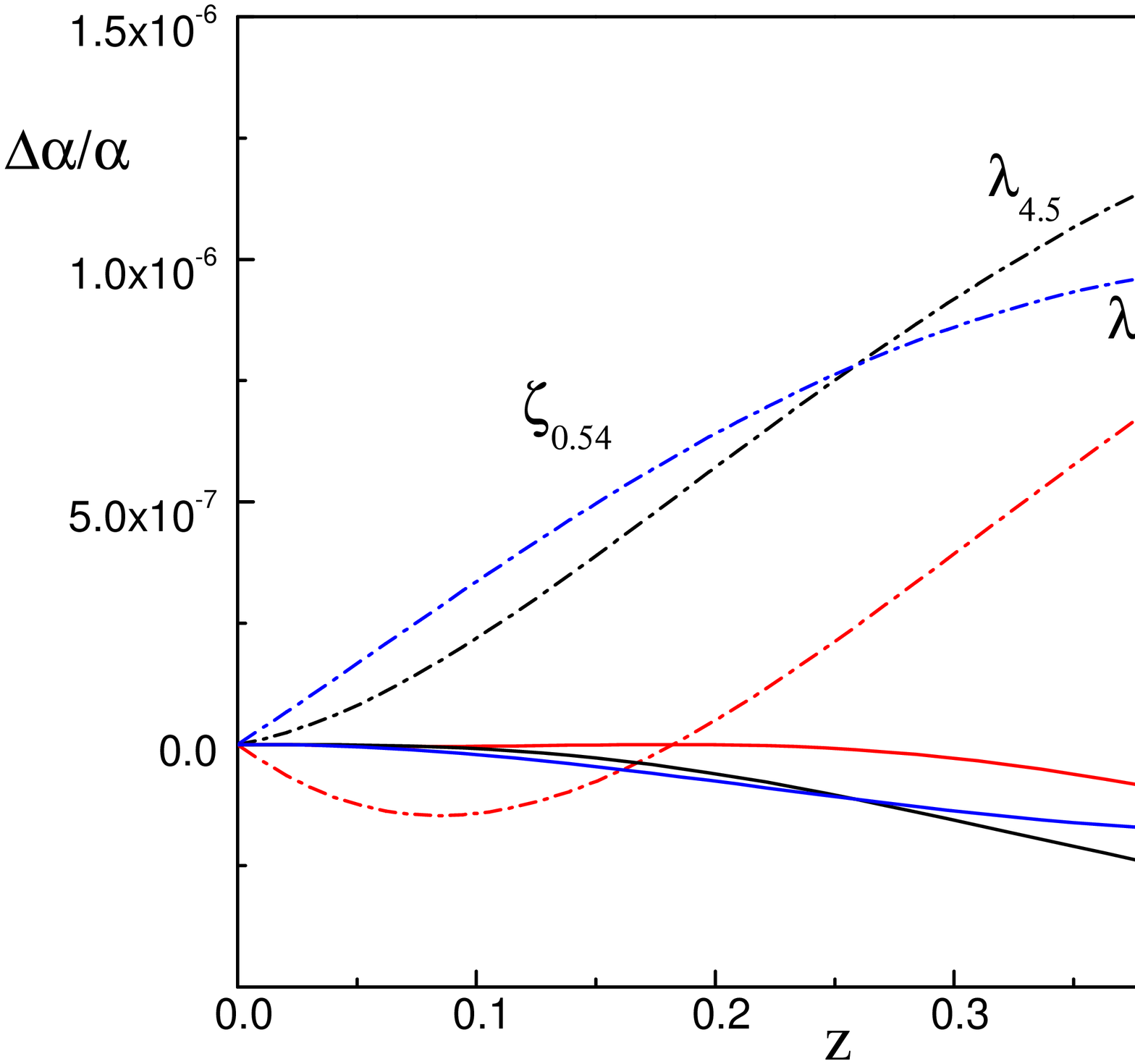, height=4.5cm}}
\epsfxsize=5cm \centerline{\psfig{file=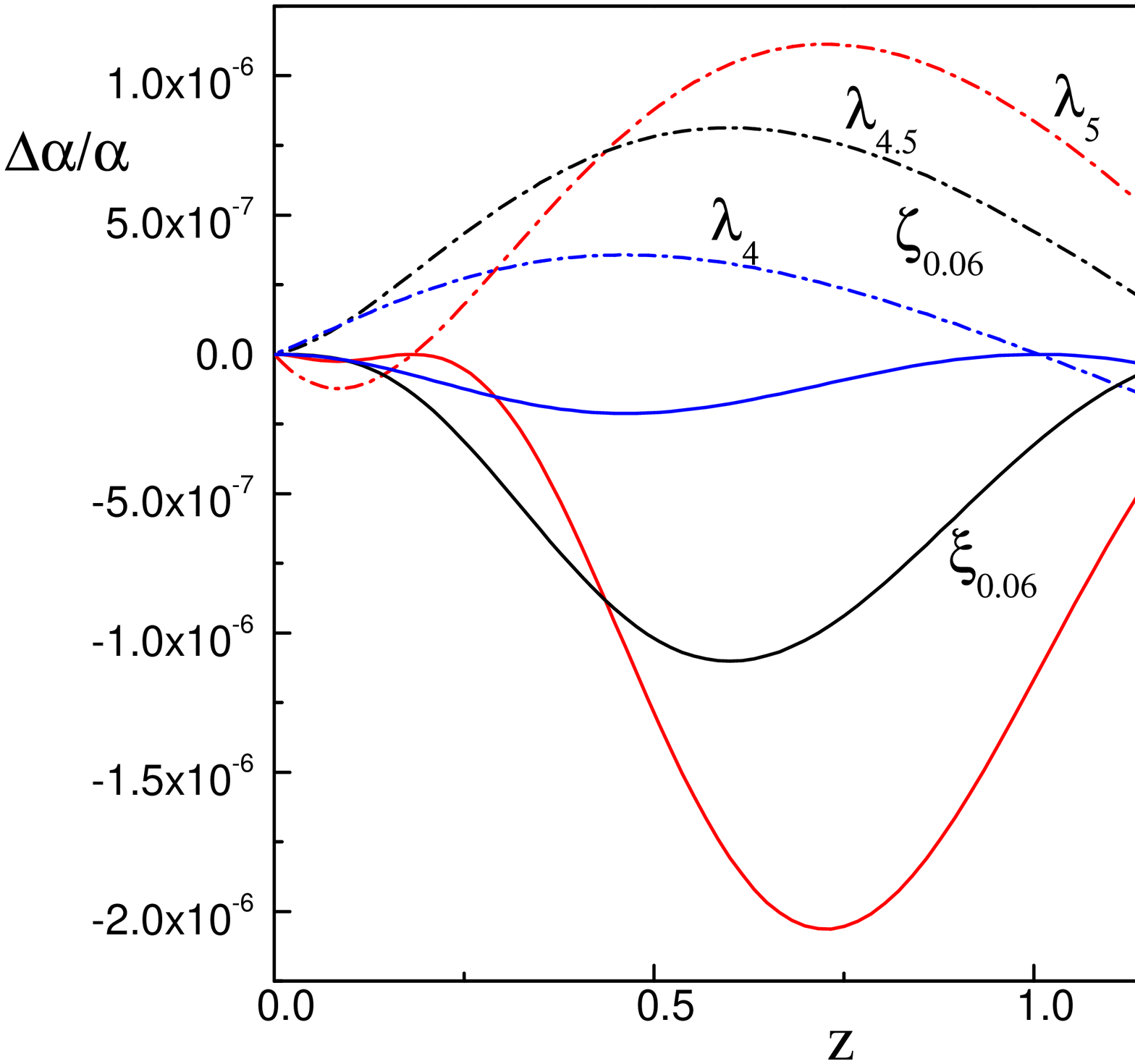,
height=4.5cm}\psfig{file=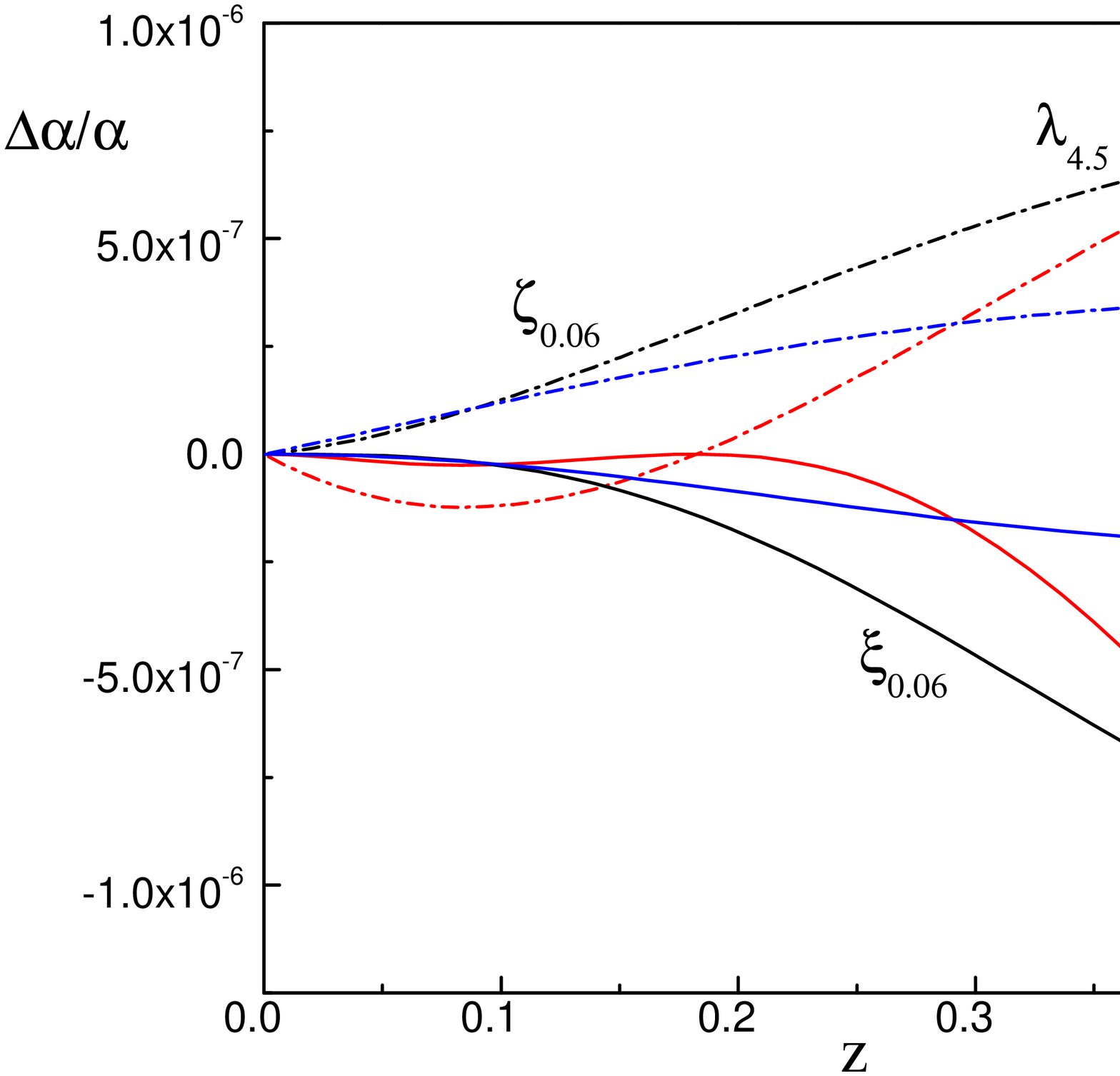, height=4.5cm}}
\vspace{-1.7cm} \caption{ As in Fig. \protect\ref{fig2} for the
potential \protect\ref{Vphi1}. Here $\lambda = 4, 4.5$, and 5. }
\label{fig3}
\end{figure}
\end{center}

\vspace{-1.0cm} In Fig.~\ref{fig2} and Fig.~\ref{fig3}, we show
the late-time evolution of $\alpha$ for the potentials considered
in the previous section. These figures are adopted from the
previous works \cite{LOP}. For the late time evolution of $\alpha$
we can use the expansion of $B_F(\phi)$ given in (\ref{dp}). In
this case, the expression for $\Delta(\alpha)$ takes the following
simple form, \be \fr{\Delta \alpha(z)}{\alpha} =
\fr{\xi}{2}\left(\phi^2(0)-\phi^2(z)\right), \label{alphaexp} \ee
where we dropped the subscript $F$ in $\xi_F$ to be concise. Using
the result of the previous section, we can predict the evolution
of $\alpha$ over redshift in terms of two parameters, $\xi$ and
$\lambda$. We choose two characteristic values of $\xi$, based on
two QSO results. To be consistent with the non-zero result for
$\Delta \alpha$ by Murphy et al. \cite{Webb}, we choose $\xi $ in
such a way that $\Delta \alpha/\alpha = -5.4 \times 10^{-6}$ at a
redshift of 3. Another option that we explore is $|\Delta
\alpha/\alpha| \leq 6\times 10^{-7}$ at $z=1.5$, which is
motivated by the experimental accuracy of Chand et al.
\cite{Petitjean}. For definiteness, in the second case we choose
$\Delta \alpha/\alpha =-7\times 10^{-6}$.

\section{Time Variation of The Proton-Electron Mass Ratio}

The proton-electron mass ratio evolves by the following simple
relation between $B_m$ and $\mu$ in our consideration, \be
\fr{\Delta \mu}{\mu} \equiv \fr{m_p(z)/m_e(z) -
m_p(0)/m_e(0)}{m_p(0)/m_e(0)} =
\fr{B_{p}(\phi(z))}{B_{e}(\phi(z))}\fr{B_{e}(\phi(0))}{B_{p}(\phi(0))}
- 1 . \label{deltamu} \ee

\begin{figure}
\vspace{1.8cm} \centerline{\psfig{file=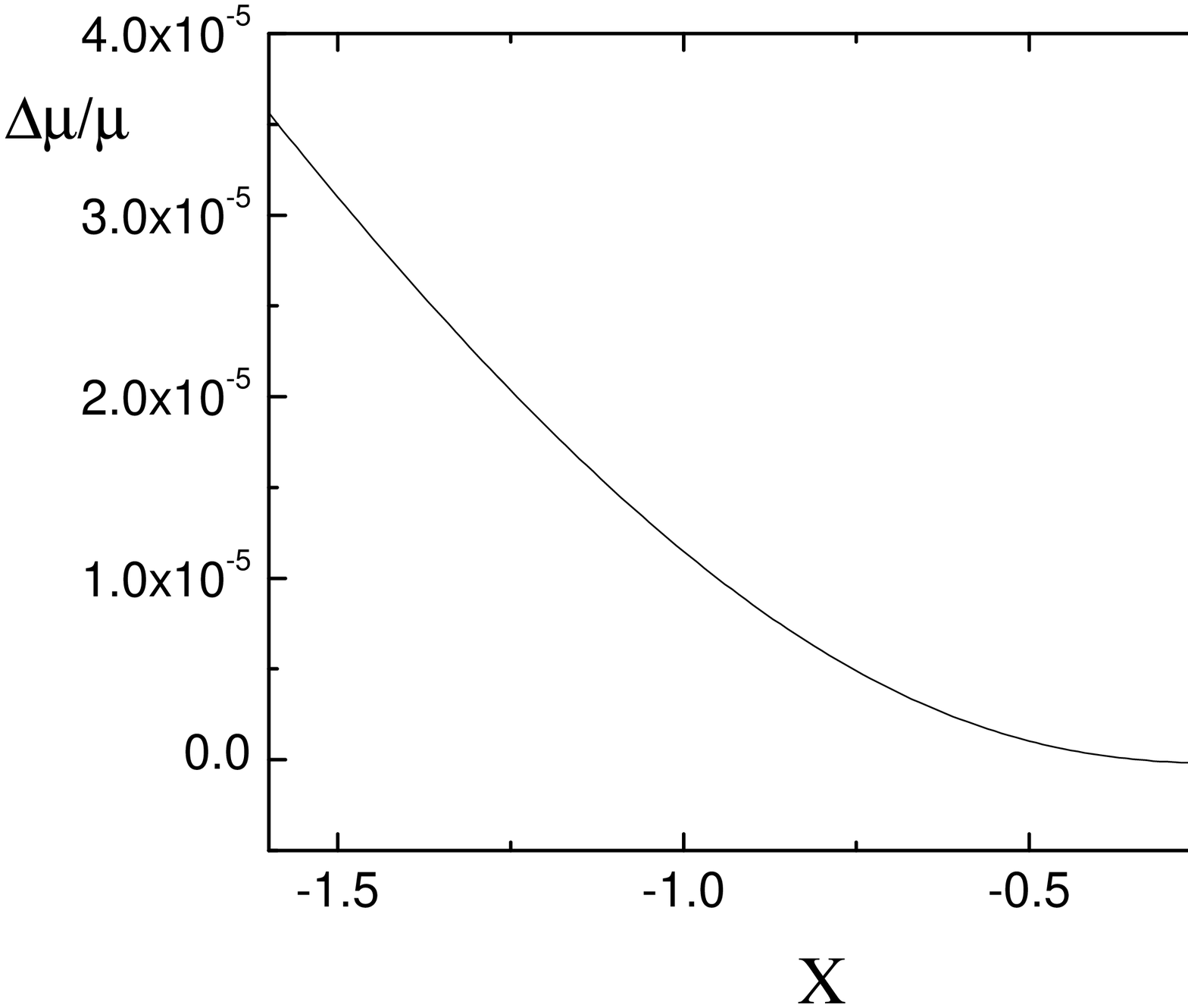,width=8cm}
\psfig{file=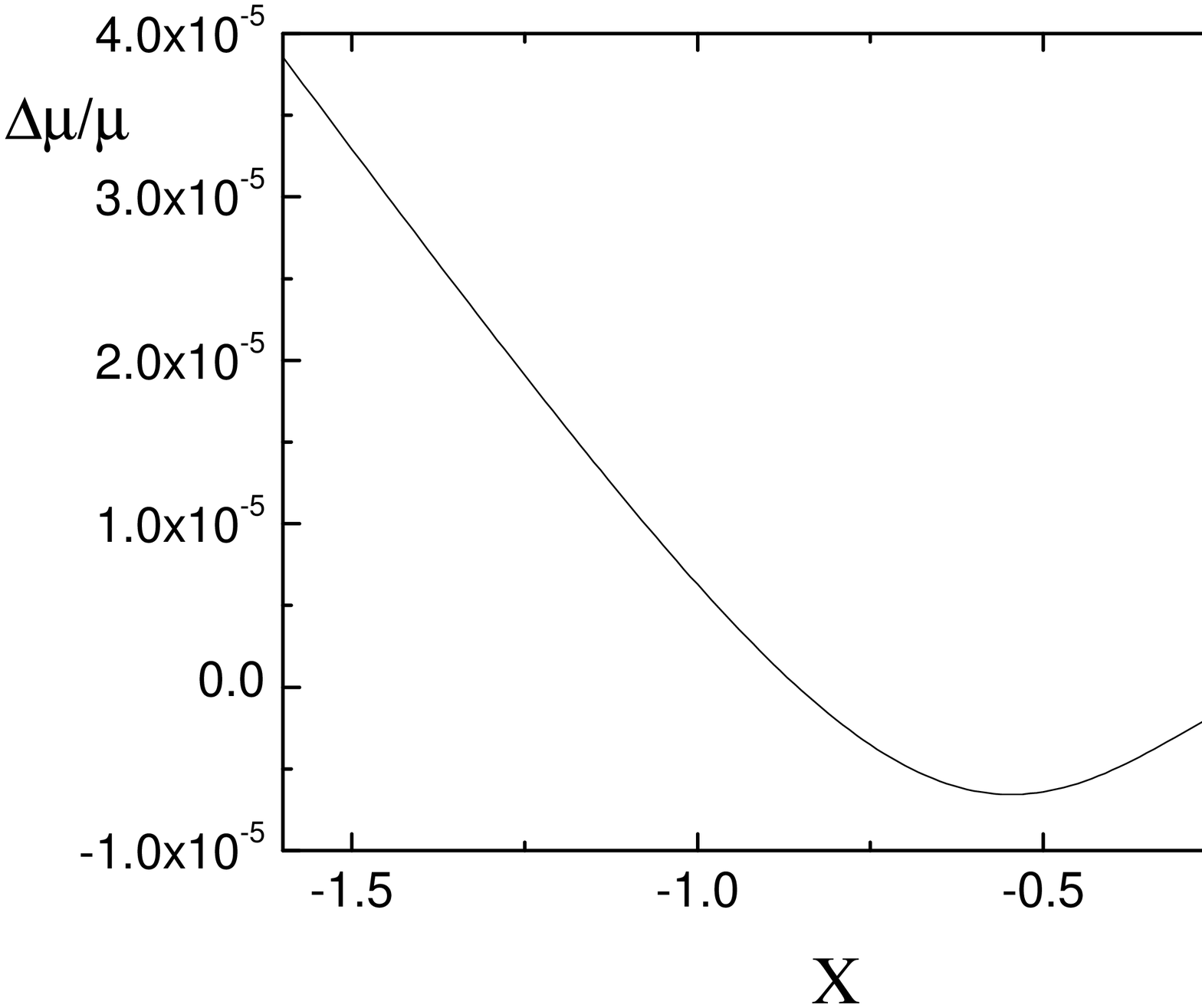,width=8cm} } \vspace{-1.7cm} \caption{The
cosmological evolution of the proton-electron mass ratio $\Delta
\mu / \mu$. a) For the $\exp(\lambda\phi^2/2)$ potential, $n = 2.5
\times 10^{-5}$. b) $n = -2.8 \times 10^{-5}$ for the
$\cosh(\lambda\phi)$ potential. \protect\label{fig4}}
\end{figure}

We can find the coupling strength $n$ for the each potential when
we adopt the result of one of the recent observation
\cite{Petitjean} around $z \sim 2.8$, \be \fr{\Delta \mu}{\mu} =
(2.4 \pm 0.6) \times 10^{-5} . \label{mm} \ee In Fig.~\ref{fig4},
we show the cosmological evolution of the proton-electron mass
ratio with the derived coupling constants $n$ for each potential.
In the left panel of the figure we show the cosmological evolution
of $\Delta \mu / \mu$ for the exponential potential. We find the
coupling constant $n = 2.5 \time 10^{-5}$ for this potential. The
right panel of the figure shows the time variation of $\Delta \mu
/ \mu$ for the $\cosh$-potential where we find $n = -2.8 \times
10^{-5}$.

\section{Conclusions}

We have analyzed the cosmological evolution of the scalar field
driven by its self-interaction potential, $V(\phi)$, and its
possible couplings to matter, $B_m(\phi)$.

We have seen that the coupling of the scalar field to the
electromagnetic field and to matters can explain the recent
observations. They can explain the time evolutions of the fine
structure constant and the proton-electron mass ratio. Our models
by use of quintessence couplings to gauge fields and matters can
pass the all the current known experimental limits.

\section*{Acknowledgments}

We thank CosPA2006 organizing committee for their hospitality and
for organizing such a nice meeting.

\end{document}